\documentclass[10pt,letterpaper]{article}
\pdfoutput=1 % if your are submitting a pdflatex (i.e. if you have
\usepackage[utf8]{inputenc} % Required for including letters with accents
\usepackage{lipsum} % Used for inserting dummy 'Lorem ipsum' text

\usepackage[top=2.5cm, bottom=3cm, left=3.5cm, right=3.5cm,
           heightrounded,marginparwidth=1.5cm, marginparsep=1cm]{geometry} %for margins

\usepackage{changepage} % for {adjustwidth} environment (margins)

\usepackage[shortlabels]{enumitem} % for lists

\usepackage[square,numbers,merge,comma,sort&compress]{natbib} % bibliography
\makeatletter
\def\NAT@spacechar{\,}  % define space inside [1,\,2] ?
\makeatother

\usepackage{amsmath,amssymb,amsfonts,amsthm,amsbsy}
\usepackage{mathtools} % for {rcases}

\usepackage{fnpct}% for multiple footnote at same point
\setfnpct{dont-mess-around} % no other kerning and punctuation switching

\usepackage{slashed,cancel}
\usepackage{old-arrows}
\usepackage{comment}   % comment environment
\usepackage{relsize}   % for \mathsmaller and \textsmaller
\usepackage{setspace}  % for double-spacing and more...
\usepackage{moresize}  % for other font sizes (\ssmall ...)
\usepackage{epsfig}
\usepackage{latexsym}
\usepackage{mathrsfs,calligra,aurical} % fonts, \mathscr ...
\usepackage{calc}
\usepackage{float}
\usepackage{appendix}
\usepackage{xargs}
\usepackage{extarrows}
\usepackage{empheq}
\usepackage{soul} % for \hl{...}

\usepackage[svgnames]{xcolor}  % Load BEFORE tikz!
\definecolor{Green}{rgb}{0.05, 0.45, 0.25}
\xdefinecolor{dogwoodrose}{rgb}{0.8, 0.1, 0.55}
\xdefinecolor{RRed}{rgb}{0.7, 0.1, 0.525}

\usepackage{tikz}
\usetikzlibrary{scopes,decorations.pathmorphing,patterns,calc,arrows,
                shapes.geometric,shapes.arrows,decorations.markings,plotmarks}
\usepackage{pgfplots}

\usepackage[blocks]{authblk}  % authors and affiliations

\usepackage[fulladjust]{marginnote}%
\usepackage{graphicx} % Required for including images
\graphicspath{{Figures/}} % Set the default folder for images

\usepackage[labelsep=colon]{caption}  % captions
\usepackage[labelsep=colon,aboveskip=10pt,belowskip=10pt]{subcaption}
\DeclareCaptionSubType * [roman]{table}
\usepackage{array,multirow,makecell,booktabs}  % tables
\newcolumntype{M}[2]{>{\centering\arraybackslash$}#1{#2\linewidth}<{$}}%centered_Maths
\newcolumntype{T}[2]{>{\centering\arraybackslash}#1{#2\linewidth}<{}}%centered_Text
\newcolumntype{R}[1]{>{\raggedleft\arraybackslash}m{#1\linewidth}<{}}
\newcolumntype{L}[1]{>{\raggedright\arraybackslash}m{#1\linewidth}<{}}
\newcommand\thinrule{\midrule[0.00001pt]}

\makeatletter
\renewcommand\mcell@classz{\@classx
   \@tempcnta \count@
   \prepnext@tok
   \@addtopreamble{%\mcell@mstyle
      \ifcase\@chnum
         \hfil
         \mcell@agape{\d@llarbegin\insert@column\d@llarend}\hfil \or
         \hskip1sp
         \mcell@agape{\d@llarbegin\insert@column\d@llarend}\hfil \or
         \hfil\hskip1sp
         \mcell@agape{\d@llarbegin \insert@column\d@llarend}\or
         \mcell@agape{$\vcenter
         \@startpbox{\@nextchar}\insert@column\@endpbox$}\or
         \mcell@agape{\vtop
         \@startpbox{\@nextchar}\insert@column\@endpbox}\or
         \mcell@agape{\vbox
         \@startpbox{\@nextchar}\insert@column\@endpbox}%
      \fi
      \global\let\mcell@left\relax\global\let\mcell@right\relax
    }\prepnext@tok}
\makeatother

\usepackage{titlesec}

\titleformat{\section}{\normalfont\fontsize{11}{11}\bfseries}{\thesection}{0.5em}{}%{\phantomsection}
\titleformat{\subsection}{\normalfont\normalsize\bfseries}{\thesubsection}{0.5em}{}%{\phantomsection}
\titleformat{\subsubsection}{\normalfont\normalsize\bfseries}{\thesubsubsection}{0.5em}{}%{\phantomsection}

\titlespacing*{\section}{0pt}%
                {4ex plus 1ex minus .5ex}{1.75ex plus .25ex minus .25ex}
\titlespacing*{\subsection}{0pt}%
                {3.5ex plus 1ex minus .5ex}{1.25ex plus .2ex minus .2ex}
\titlespacing*{\subsubsection}{0pt}%
                {2.5ex plus 0.75ex minus .2ex}{0.75ex plus .15ex minus .15ex}
\titlespacing*{\paragraph}{0pt}%
                {1.85ex plus 0.5ex minus .15ex}{1em}

\usepackage{titletoc}

\addtocontents{toc}{\addvspace{-0.75em}}  % reduce vert space before TOC

\titlecontents{section}
  [1.25em] {\addvspace{0.7em plus 0pt}\small}
  {\thecontentslabel\hspace{0.75em}}{}%{\thecontentslabel\hspace{0.75em}}
  {\hspace{0.5em}\titlerule*[0.5em]{.}\contentspage}
  [\addvspace{0.0em plus 0pt}]

\titlecontents{subsection}
  [2.75em] {\addvspace{0.075em plus0pt}\fns}
  {\thecontentslabel\hspace{0.75em}}{\thecontentslabel\hspace{0.75em}}
  {\hspace{0.5em}\titlerule*[0.5em]{.}\small\contentspage}
  [\addvspace{0.075em plus 0pt}]

\usepackage{environ}
\makeatletter
\NewEnviron{subalign}[1]{
\begin{subequations}\label{#1}
\begin{align} \BODY \end{align}
\end{subequations}      }
\makeatother
\makeatletter
\newenvironment{subeqs}%
{\begingroup%
\setlength{\abovedisplayskip}{10pt plus 4pt minus 9pt}%
\setlength{\abovedisplayshortskip}{0pt plus 2pt minus 2pt}%
\setlength{\belowdisplayskip}{12pt plus 3pt minus 9pt}%
\setlength{\belowdisplayshortskip}{7pt plus 3pt minus 4pt}%
\begin{subequations}%
}%
{\end{subequations}\ignorespacesafterend%
\endgroup}%
\makeatother

\makeatletter
{\begingroup%
\setlength{\abovedisplayskip}{{#1}pt plus 3pt minus 3pt}%
\setlength{\abovedisplayshortskip}{{#2}pt plus 3pt}%
\setlength{\belowdisplayskip}{{#3}pt plus 3pt minus 3pt}%
\setlength{\belowdisplayshortskip}{{#4}pt plus 3pt minus 3pt}%
\begin{subequations}%
}%
{\end{subequations}\ignorespacesafterend%
\endgroup}%
\makeatother

\makeatletter
{\begingroup%
\setlength{\abovedisplayskip}{{#1}pt plus 3pt minus 3pt}%
\setlength{\abovedisplayshortskip}{{#2}pt plus 3pt}%
\setlength{\belowdisplayskip}{{#3}pt plus 3pt minus 3pt}%
\setlength{\belowdisplayshortskip}{{#4}pt plus 3pt minus 3pt}%
\begin{equation}%
}%
{\end{equation}\ignorespacesafterend%
\endgroup}%
\makeatother

\makeatletter
{%
\begin{equation}%
\begin{split}%
}%
{\end{split}%
\end{equation}\ignorespacesafterend%
}%
\makeatother

\usepackage{bm}  % \mathbbm only has digits "1" and "2"
\usepackage{dsfont}  % double math characters (but only digit "1")

\DeclareMathAlphabet{\mathpzc}{OT1}{pzc}{m}{it}
\DeclareMathAlphabet{\mathcal}{OMS}{cmsy}{m}{n}
\DeclareSymbolFontAlphabet{\Scr}{rsfs}
\DeclareMathAlphabet{\mathbold}{U}{BOONDOX-ds}{m}{n}
\SetMathAlphabet{\mathbold}{bold}{U}{BOONDOX-ds}{b}{n}
\DeclareMathAlphabet{\mathcalboondox}{U}{BOONDOX-calo}{m}{n}
\SetMathAlphabet{\mathcalboondox}{bold}{U}{BOONDOX-calo}{b}{n}
\DeclareMathAlphabet{\mathbcalboondox}{U}{BOONDOX-calo}{b}{n}

\newcommand\linkcol{RRed}


\usepackage[breaklinks=true,backref=page]{hyperref}
\hypersetup{
    bookmarks=true,         % show bookmarks bar?
    bookmarksnumbered=true,
    pdftoolbar=true,        % show Acrobat’s toolbar?
    pdfmenubar=true,        % show Acrobat’s menu?
    pdffitwindow=false,     % window fit to page when opened
    pdfauthor={},     % author
    pdfsubject={},   % subject of the document
    pdfcreator={},   % creator of the document
    pdfproducer={},  % producer of the document
    pdfpagemode={UseNone},
    pdfstartview={FitH},
    colorlinks=true,
    plainpages,
    linktoc=page,
    citecolor=blue,
    filecolor=black,
    linkcolor=\linkcol,
    urlcolor=Green,
}
\renewcommand*{\backref}[1]{}
\renewcommand*{\backrefalt}[4]{%
\ifcase #1 %
\relax
\or
~{\small [\textsc{p.~\fns{\!#2}}]}
\else
~{\small [\textsc{p.~\fns{\!#2}}]}%
\fi}

\usepackage{footnotebackref}
\usepackage{hypernat} % Load it HERE, after NATBIB and HYPERREF
\usepackage{cleveref} % Load it HERE, after HYPERREF

\def\+{~+~}
\def\-{~-~}
\def\={\:=\:}
\newcommand\fns{\footnotesize}

\newcommand\qRq{\quad\Rightarrow\quad}

\newcommand\Real{\textrm{Re}}
\newcommand\Img{\textrm{Im}}

\newcommand\eps{\varepsilon}
\newcommand\epsz{\varepsilon_\ms{0}}
\newcommand\epsg{\varepsilon_\textrm{g}}
\newcommand\mug{\mu_\textrm{g}}
\newcommand\muz{\mu_\ms{0}}
\newcommand\w{\omega}

\newcommand\hgamma{\bar{\gamma}}

\newcommand\Tc{T_\textrm{c}}
\newcommand\ux{\vec{u}_x}
\newcommand\E{\mathbf{E}}
\newcommand\B{\mathbf{B}}
\newcommand\A{\mathbf{A}}

\newcommand\Bc{B_\textrm{c}}

\newcommand\Eg{\mathbf{E}_\textrm{g}}
\newcommand\Bg{\mathbf{B}_\textrm{g}}
\newcommand\Ag{\mathbf{A}_\textrm{g}}
\newcommand\Ee{\mathbf{E}_\textrm{e}}
\newcommand\Be{\mathbf{B}_\textrm{e}}
\newcommand\Ae{\mathbf{A}_\textrm{e}}
\newcommand\jj{\mathbf{j}}
\newcommand\jg{\mathbf{j}_\textrm{g}}

\newcommand\x{\mathbf{x}}

\newcommand\gstar{g_\star}

\newcommand\rhog{\rho_\textrm{g}}
\newcommand\phig{\phi_\textrm{g}}
\newcommand\vF{v_\textsc{f}}

\newcommand\mt{\mathrm{m}}

\newcommand\s{\mathrm{s}}
\newcommand\Kelv{\mathrm{K}}
\newcommand\GN{\mathrm{G}}%{_\ms{\textsc{n}}}

\providecommand{\abs}[1]{\left\lvert#1\right\rvert}

\newcommand{\ms}{\mathsmaller}

\newcommand{\dd}{\partial}
\newcommandx{\tcR}[1]{\textcolor{Crimson}{#1}}
\newcommandx{\tts}[1]{\text{\textsmaller{#1}}}
\newcommandx{\dt}[1][1=f,usedefault]{\frac{\partial{#1}}{\partial t}}
\newcommandx{\dtau}[1][1=f,usedefault]{\frac{\partial{#1}}{\partial\tau}}
\newcommandx{\dx}[1][1=f,usedefault]{\frac{\partial{#1}}{\partial x}}
\newcommandx{\ddx}[1][1=f,usedefault]{\frac{\partial^2{#1}}{{\partial x}^2}}
\newcommandx{\dm}[1][1=\mu,usedefault]{\partial_{#1}}
\newcommandx{\dmup}[1][1=\mu,usedefault]{\partial^{#1}}
\newcommandx{\subm}[2][1=p,2=A,usedefault]{{#1}_{\!\mathsmaller{#2}}}
\newcommandx{\subt}[2][1=p,2=A,usedefault]{{#1}_\text{\textsmaller{#2}}}
\newcommandx{\supm}[2][1=p,2=A,usedefault]{{#1}^{\!\mathsmaller{#2}}}
\newcommandx{\supt}[2][1=p,2=A,usedefault]{{#1}^\text{\textsmaller{#2}}}
\newcommandx{\subpt}[3][1=p,2=A,3=B,usedefault]{{#1}^\text{\textsmaller{#3}}_\text{\textsmaller{#2}}}
\newcommandx{\subpm}[3][1=p,2=A,3=B,usedefault]{{#1}^{\mathsmaller{#3}}_{\mathsmaller{#2}}}
\newcommandx{\sh}[1][1=\alpha,usedefault]{\sinh\left(#1\right)}
\newcommandx{\ch}[1][1=\alpha,usedefault]{\cosh\left(#1\right)}
\newcommandx{\sech}[1][1=\alpha,usedefault]{\mathrm{sech}\left(#1\right)}
\newcommandx{\cosech}[1][1=\alpha,usedefault]{\mathrm{cosech}\left(#1\right)} \newcommandx{\LCTd}[4][1=\mu,2=\nu,3=\rho,4=\sigma,usedefault]{\eps_{#1#2#3#4}}
\newcommandx{\LCTu}[4][1=\mu,2=\nu,3=\rho,4=\sigma,usedefault]{\eps^{#1#2#3#4}}

\newcommandx{\gmetr}[2][1=\mu,2=\nu,usedefault]{g_{{#1}{#2}}}
\newcommandx{\invgmetr}[2][1=\mu,2=\nu,usedefault]{g^{{#1}{#2}}}
\newcommandx{\spc}[3][1=\mu,2=a,3=b,usedefault]{{\w_{#1}}^{\!\!{#2}{#3}}}
\newcommandx{\Conn}[3][1=\mu,2=\nu,3=\lambda,usedefault]{{\Gamma_{{#1}{#2}}}^{\!\!#3}}
\newcommandx{\viel}[2][1=\mu,2=a,usedefault]{{e_{#1}}^{\!#2}}
\newcommandx{\inviel}[2][1=a,2=\mu,usedefault]{{e_{#1}}^{#2}}
\newcommandx{\vieluu}[2][1=\mu,2=a,usedefault]{e^{#1#2}}
\newcommandx{\Rdduu}[4][1=\mu,2=\nu,3=a,4=b,usedefault]{{R_{{#1}{#2}}}^{{#3}{#4}}}
\newcommandx{\hgamui}[1][1=0,usedefault]{\hgamma^{\mathsmaller{#1}}}
\newcommandx{\hgamdi}[1][1=0,usedefault]{\hgamma_{{}_{#1}}}
\newcommandx{\gamui}[1][1=0,usedefault]{\gamma^{\mathsmaller{#1}}}
\newcommandx{\gamdi}[1][1=0,usedefault]{\gamma_{{}_{#1}}}

\newcommandx{\emetr}[2][1=\mu,2=\nu,usedefault]{\eta_{{#1}{#2}}}
\newcommandx{\invemetr}[2][1=\mu,2=\nu,usedefault]{\eta^{{#1}{#2}}}
\newcommandx{\hmetr}[2][1=\mu,2=\nu,usedefault]{h_{{#1}{#2}}}
\newcommandx{\invhmetr}[2][1=\mu,2=\nu,usedefault]{h^{{#1}{#2}}}
\newcommandx{\bhmetr}[2][1=\mu,2=\nu,usedefault]{\bar{h}_{{#1}{#2}}}
\newcommandx{\binvhmetr}[2][1=\mu,2=\nu,usedefault]{\bar{h}^{{#1}{#2}}}
\newcommandx{\hud}[2][1=\mu,2=\nu,usedefault]{{h^{#1}}_{\!\!#2}}
\newcommandx{\Ruddd}[4][1=\sigma,2=\mu,3=\lambda,4=\nu,usedefault]{{R^{#1}}_{\!{#2}{#3}{#4}}}
\newcommandx{\Gam}[3][1=\lambda,2=\mu,3=\nu,usedefault]{{\Gamma^{#1}}_{\!{#2}{#3}}}
\newcommandx{\Gamd}[3][1=\mu,2=\nu,3=\lambda,usedefault]{\Gamma_{{#1}{#2}{#3}}}
\newcommandx{\Ricci}[2][1=\mu,2=\nu,usedefault]{R_{{#1}{#2}}}
\newcommandx{\GEinst}[2][1=\mu,2=\nu,usedefault]{G^{{}^\tts{(E)}}_{{#1}{#2}}}
\newcommandx{\Gscr}[3][1=\mu,2=\nu,3=\rho,usedefault]{\mathscr{G}_{{#1}{#2}{#3}}}
\makeatletter
\normalsize
\divide\@tempdima\baselineskip
\@tempcnta=\@tempdima
\makeatother

\DeclareFixedFont\trfont{OT1}{phv}{b}{sc}{11}

\title{%
       \vspace{-1.0cm}
       \centering\boldmath\LARGE\bfseries%
       Possible alterations of local gravitational field inside a superconductor
       \bigskip
       }

\author{\textsc{Giovanni Alberto Ummarino}
\vspace{0.1em}}
\affil{%
{Politecnico di Torino, Dipartimento di Scienza Applicata e Tecnologia, corso Duca degli Abruzzi 24, 10129 Torino, Italy}%
}
\affil{National Research Nuclear University MEPhI, Kashirskoe hwy 31, 115409
       Moscow, Russia%
\vspace{-0.025em}
}%
\affil{\href{mailto:giovanni.ummarino@polito.it}{\texttt{giovanni.ummarino@polito.it}}
       }

\smallskip

\author{\textsc{Antonio Gallerati}
\vspace{0.1em}
}
\affil{%
{Politecnico di Torino, Dipartimento di Scienza Applicata e Tecnologia, corso Duca degli Abruzzi 24, 10129 Torino, Italy}%       %{\,\raisebox{-0.1\baselineskip}{\protect\includegraphics[height=1em]{polito_logo.jpg}}}
       }
\affil{Istituto Nazionale di Fisica Nucleare, Sezione di Torino, via Pietro
       Giuria 1, 10125 Torino, Italy%
\vspace{-0.025em}
       }%
\affil{\href{mailto:antonio.gallerati@polito.it}{\texttt{antonio.gallerati@polito.it}}
      }

\date{}

\makeatletter
\patchcmd{\@maketitle}{\begin{center}}{\begin{adjustwidth}{-0.25in}{-0.25in}\begin{center}}{}{}
\patchcmd{\@maketitle}{\end{center}}{\end{center}\end{adjustwidth}}{}{}
\makeatother

\begin{document}

\maketitle
\smallskip

\begin{abstract}
\noindent
We calculate the possible interaction between a superconductor and the static Earth's gravitational fields, making use of the gravito-Maxwell formalism combined with the time-dependent Ginzburg--Landau theory.
We try to estimate which are the most favourable conditions to enhance the effect, optimizing the superconductor parameters characterizing the chosen sample. We also give a qualitative comparison of the behaviour of high--$\Tc$ and classical low--$\Tc$ superconductors with respect to the gravity/superfluid interplay.
\end{abstract}

\bigskip

\tableofcontents
%\listoffigures
%\listoftables
\pagebreak

%\bigskip

%%%%%%%%%%%%%%%%%%%%%%%%%%%%%%%

\section{Introduction} \label{sec:Intro}
The study of possible gravitational effects on superconductors is more than 50 years old and started with the seminal paper of DeWitt \cite{DeWitt:1966yi}. In the following years, there has been a fair amount of scientific literature on the subject
\cite{papini1967detection,Papini:1970cw,rothen1968application,rystephanick1973london,hirakawa1975superconductors,minasyan1976londons,anandan1977gravitational,Anandan:1978na,anandan1984relthe,anandan1994relgra,ross1983london,Felch:1985pre,dinariev1987relativistic,peng1990new,peng1991electrodynamics,peng1991interaction,li1991effects,li1992gravitational,torr1993gravitoelectric,de1992torsion},
but it was only after the 1992 Podkletnov's reported effect \cite{podkletnov1992possibility,podkletnov1997weak} that experimental, laboratory configurations were proposed to detect the interaction.\par
Theoretical interpretations of the interplay between the condensate and the local gravitational field were produced in 1996 exploiting the framework of quantum gravity \cite{Modanese:1995tx}, showing how a suitable Lagrangian coupling of the superfluid can determine a gravitational interaction with the condensate and consequent localized slight instabilities \cite{modanese1996role,Wu:2003aq}. Although being a solid and elegant formulation offering a general, theoretical explanation for the described interplay, the quantum gravity approach involves a formalism that makes it hard to extract quantitative predictions.\par\smallskip
Parallel to DeWitt (and related) studies about gravity/supercondensate coupling,  other theoretical \cite{schiff1966gravitation,Dessler:1968zz} and experimental \cite{witteborn1967experimental,witteborn1968experiments,herring1968gravitationally} researches were conducted about electric-type fields induced in conductors by the presence of the gravitational field, analysing the importance of the internal structure of special classes of solids and fluids when gravity is taken into account. Those researches also inspired other recent papers that focus on various relevant aspects of the behaviour of superconductors interacting with gravitational waves \cite{Minter:2009fx,Quach:2015qwa,quach2020fisher}.\par
One of the results of the above studies was the introduction of a fundamental, generalized electric-like field, featuring an electrical component and a gravitational one. In the following, we are going to extend those results making use of the \emph{gravito-Maxwell formalism} \cite{Ummarino:2017bvz,Behera:2017voq,Ummarino:2019cvw,Ummarino:2020loo,Gallerati:2020tyq}. In particular, we will see that the latter approach can provide a solid framework where to obtain a generalized form for the electric/magnetic fields, involved in quantum effects originating from the interaction with the weak gravitational background. On the other side, the formalism also turns out to be powerful in the study of gravity/superconductivity interplay, since the formal analogy between the Maxwell and weak gravity equations allows us to use the Ginzburg--Landau theory for the microscopic description of the interaction. We will in fact analyse how the weak local gravitational field can be affected by the presence of the superfluid condensate, writing explicit time-dependent Ginzburg--Landau equations for the superconductor order parameter.\par
With respect to our previous analysis \cite{Ummarino:2017bvz}, we will perform new calculations in a different gauge and this will lead us to clearer and deeper conclusions on the interpretation of the conjectured effect. We will also analyse which parameters could be optimized to enhance the interaction, choosing appropriate conditions and sample characteristics.

%%%%%%%%%%%%%%%%%%%%%%%%%%%%%%%%%%%%%%%%%%%%%%%%%%%%%%%

\section{Generalized gravito-Maxwell equations}%
\label{sec:gravMax}
\sloppy
Let us consider a nearly--flat spacetime configuration (weak, static gravitational field approximation) so that the metric can be expanded as:
\begin{equation} \label{eq:gmetr}
\gmetr~\simeq~\emetr+\hmetr\;,
\end{equation}
where the symmetric tensor $\hmetr$ is a small perturbation of the constant, flat Minkowski metric in the mostly plus convention, $\emetr=\mathrm{diag}(-1,+1,+1,+1)$. The inverse metric, in linear approximation, is given by
\begin{equation}
\invgmetr~\simeq~\invemetr-\invhmetr\;.
\end{equation}
while the metric determinant can be expanded as
\begin{align}
g\,=\,\det\left[\gmetr\right]
    \,=\,\varepsilon^{\mu\nu\rho\sigma}\gmetr[1][\mu]\,\gmetr[2][\nu]\,\gmetr[3][\rho]\,\gmetr[4][\sigma]
    \,\simeq\,-1-h
\;\quad\Rightarrow\quad\;
\sqrt{-g}\:\simeq\,1+\frac12\,h\;,\qquad\qquad
\end{align}
where $h=\hud[\sigma][\sigma]$.

\subsection{Generalizing Maxwell equations}
If we consider an inertial coordinate system, to linear order in $\hmetr$ the connection is expanded as
\begin{equation}\label{eq:Gam}
\Gam[\lambda][\mu][\nu]~\simeq~\frac12\,\invemetr[\lambda][\rho]\,
     \left(\dm[\mu]\hmetr[\nu][\rho]+\dm[\nu]\hmetr[\rho][\mu]
     -\dm[\rho]\hmetr[\mu][\nu]\right)\:.
\end{equation}
The Riemann tensor is defined as:
\begin{equation}
\begin{split}
\Ruddd&\=\dm[\lambda]\Gam[\sigma][\mu][\nu]-\dm[\nu]\Gam[\sigma][\mu][\lambda]
        +\Gam[\sigma][\rho][\lambda]\,\Gam[\rho][\nu][\mu]
        -\Gam[\sigma][\rho][\nu]\,\Gam[\rho][\lambda][\mu]\:,
%\\
%    &\= 2\,\dm[{[}\lambda]\Gam[\sigma][\nu{]}][\mu]
%    \+2\,\Gam[\sigma][\rho][{[}\lambda]\,\Gam[\rho][\nu{]}][\mu]\;,
\end{split}
\end{equation}
while the Ricci tensor is given by the contraction
\begin{equation}
\Ricci\=\Ruddd[\sigma][\mu][\sigma][\nu]\:,
\end{equation}
and, to linear order in $\hmetr$, it reads
\begin{equation} \label{eq:Ricci}
\begin{split}
\Ricci&\:\simeq\:
    \dm[\sigma]\Gam[\sigma][\mu][\nu]+\dm[\mu]\Gam[\sigma][\sigma][\nu]+\cancel{\Gamma\,\Gamma}-\cancel{\Gamma\,\Gamma}
    \=\frac12\,\left(\dm\dmup[\rho]\hmetr[\nu][\rho]+\dm[\nu]\dmup[\rho]\hmetr[\mu][\rho]\right)
      -\frac12\,\dm[\rho]\dmup[\rho]\hmetr-\frac12\,\dm\dm[\nu]h\=
\\[\jot]
   &\=\dmup[\rho]\dm[{(}\mu]\hmetr[\nu{)}][\rho]-\frac12\,\dd^2\hmetr-\frac12\,\dm\dm[\nu]h\:,
\end{split}
\end{equation}
having used eq.\ \eqref{eq:Gam}.\par\smallskip
The Einstein equations have the form \cite{Wald:1984rg}:
\begin{equation}
\Ricci-\dfrac12\,\gmetr\,R\=8\pi\GN\;T_{\mu\nu}\:,
\label{eq:Einstein}
\end{equation}
where $R=\invgmetr\Ricci$ is the Ricci scalar. In first-order approximation, we can write
\begin{equation}
\frac12\,\gmetr\,R~\simeq~
    \frac12\,\emetr\,\invemetr[\rho][\sigma]\Ricci[\rho][\sigma]
    \=\frac12\,\emetr\,\left(\dmup[\rho]\dmup[\sigma]\hmetr[\rho][\sigma]
       -\dd^2h\right)\:,
\end{equation}
having used eq.\ \eqref{eq:Ricci}, and the left hand side of \eqref{eq:Einstein} turns out to be
\begin{equation}
\begin{split}
\Ricci-\dfrac12\,\gmetr\,R~\simeq~
    \dmup[\rho]\dm[{(}\mu]\hmetr[\nu{)}][\rho]-\frac12\,\dd^2\hmetr-\frac12\,\dm\dm[\nu]h
    -\frac12\,\emetr\left(\dmup[\rho]\dmup[\sigma]\hmetr[\rho][\sigma]-\dd^2h\right)\:.
\label{eq:lhsEinstein}
\end{split}
\end{equation}
Now, we introduce the symmetric traceless tensor
\begin{equation}
\bhmetr\=\hmetr-\frac12\,\emetr\,h\:,
\end{equation}
so that the above \eqref{eq:lhsEinstein} can be rewritten as
\begin{equation}
\begin{split}
\Ricci-\dfrac12\,\gmetr\,R~\simeq~
    &\frac12\left(\dmup[\rho]\dm\bhmetr[\nu][\rho]+\dmup[\rho]\dm[\nu]\bhmetr[\mu][\rho]
     -\dmup[\rho]\dm[\rho]\bhmetr[\mu][\nu]
     -\emetr\,\dmup[\rho]\dmup[\sigma]\bhmetr[\rho][\sigma]\right)%\=
%\\
=~\dmup[\rho]\dm[{[}\nu]\bhmetr[\rho{]}][\mu]+\dmup[\rho]\dmup[\sigma]\emetr[\mu][{[}\sigma]\,\bhmetr[\nu{]}][\rho]\=\qquad
\\[1.5\jot]
=~&\dmup[\rho]\left(
    %\underbrace{
         \dm[{[}\nu]\bhmetr[\rho{]}][\mu]+\dmup[\sigma]\emetr[\mu][{[}\rho]\,\bhmetr[\nu{]}][\sigma]
    %}_{\mathclap{\Gscr}}
         \right)\:.
%~\equiv~\dmup[\rho]\,\Gscr\;,
\end{split}
\end{equation}
We then define the tensor
\begin{equation} \label{eq:Gscr}
\Gscr~\equiv~
\dm[{[}\nu]\bhmetr[\rho{]}][\mu]+\dmup[\sigma]\emetr[\mu][{[}\rho]\,\bhmetr[\nu{]}][\sigma]\:,
%\qquad\qquad \left(\Gscr=-\Gscr[\mu][\rho][\nu]\right)\:,
\end{equation}
so that the Einstein equations can be finally recast in the compact form:
\begin{equation}
\dmup[\rho]\Gscr\=8\pi\GN\;T_{\mu\nu}\:.
\label{eq:Einstein_dG}
\end{equation}
\par\smallskip

\paragraph{Gauge fixing.}
We now consider the \emph{harmonic coordinate condition}, expressed by the relation \cite{Wald:1984rg}:
\begin{equation}
\dm\left(\sqrt{-g}\,\invgmetr\right)=0
\;\quad\Leftrightarrow\quad\;
\Box x^\mu=0\:,
\label{eq:harmcond}
\end{equation}
that in turn can be rewritten in the form
\begin{equation}
\invgmetr\,\Gam\,=\,0\:,
\label{eq:DeDonder}
\end{equation}
also known as \emph{De Donder gauge}. The requirement of the above coordinate condition \eqref{eq:harmcond} plays then the role of a gauge fixing. %
%removing indeterminacy and determining the physical configuration
%\footnote{%
%in particular, in harmonic coordinates the metric satisfies a manifestly Lorenz-covariant condition, so that the De Donder gauge becomes a natural choice. Moreover, if one considers the weak-field expansion of the Einstein-Hilbert action in De Donder gauge, the action itself (as well as the graviton propagator) takes a particularly simple form.}
Imposing the above \eqref{eq:DeDonder} and using eqs.\ \eqref{eq:gmetr} and
\eqref{eq:Gam}, in linear approximation we find:
\begin{equation}
0\:\simeq\: \frac12\,\invemetr\,\invemetr[\lambda][\rho]\left(\dm[\mu]\hmetr[\nu][\rho]+\dm[\nu]\hmetr[\rho][\mu]-\dm[\rho]\hmetr[\mu][\nu]\right)\=
\dm\invhmetr[\mu][\lambda]-\frac12\,\dmup[\lambda]h\:,
\end{equation}
that is, we have the condition
\begin{equation} \label{eq:gaugecond0}
\dm\invhmetr\simeq\frac12\,\dmup[\nu]h
\;\quad\Leftrightarrow\;\quad
\dmup\hmetr\simeq\frac12\,\dm[\nu]h\;.
\end{equation}
Now, one also has
\begin{equation}
\dmup\hmetr\=\dmup\left(\bhmetr+\frac12\,\emetr h\right)
\=\dmup\bhmetr+\frac12\,\dm[\nu]h\:,
\end{equation}
and, using eq.\ \eqref{eq:gaugecond0}, we find the so-called \emph{Lorentz gauge condition}:
\begin{equation}
\dmup\bhmetr\:\simeq\:0\:.
\label{eq:Lorentzgauge}
\end{equation}
The above relation further simplifies expression \eqref{eq:Gscr} for $\Gscr$, which takes the very simple form
\begin{equation}
\Gscr~\simeq~\dm[{[}\nu]\bhmetr[\rho{]}][\mu]\:,
\end{equation}
and verifies also the relation
\begin{equation}
\dm[{[}\lambda{|}]\Gscr[0][{|}\mu][\nu{]}]\=0
\;\qRq\;
\Gscr[0][\mu][\nu] \propto \dm\mathcal{A}_\nu-\dm[\nu]\mathcal{A}_\mu\:,
\end{equation}
implying the existence of a potential (see next paragraph).
\par\medskip

\paragraph{Gravito-Maxwell equations.}
Now, let us define the fields%
\footnote{for the sake of simplicity, we initially set the physical charge $e=m=1$}
\begin{subeqs}\label{eq:fields0}
\begin{align}
\Eg&~\equiv~E_i~=\,-\,\frac12\,\Gscr[0][0][i]~=\,-\,\frac12\,\dm[{[}0]\bhmetr[i{]}][0]\:,\\[\jot]
\Ag&~\equiv~A_i\=\frac14\,\bhmetr[0][i]\:,\\[\jot]
\Bg&~\equiv~B_i%\=\frac12\left(\Gscr[0][2][3],\,\Gscr[0][3][1],\,\Gscr[0][1][2]\right)
                 \=\frac14\,{\varepsilon_i}^{jk}\,\Gscr[0][j][k]\:,
\end{align}
\end{subeqs}%   always put % here!
where $i=1,2,3$ and
\begin{equation}
\Gscr[0][i][j]\=\dm[{[}i]\bhmetr[j{]}][0]
\=\frac12\left(\dm[i]\bhmetr[j][0]-\dm[j]\bhmetr[i][0]\right)\=4\,\dm[{[}i]A_{j{]}}\:.
\end{equation}
One can immediately see that
\begin{equation}
\begin{split}
\Bg&\=\frac14\,{\varepsilon_i}^{jk}\,4\,\dm[{[}j]A_{k{]}}
   \={\varepsilon_i}^{jk}\,\dm[j]A_k=\nabla\times\Ag\:,\\[3\jot]
%&\Longrightarrow\quad \Bg=\nabla\times\Ag\;,%\\[3\jot]
&~\Longrightarrow\quad \nabla\cdot\Bg\=0\:.
\end{split}
\end{equation}
%\begin{equation}
%\Bg\=\frac14\,{\varepsilon_i}^{jk}\,4\,\dm[{[}j]A_{k{]}}
%   \={\varepsilon_i}^{jk}\,\dm[j]A_k=\nabla\times\Ag\;,%\;,\\[3\jot]
%\end{equation}
%%
%that also implies
%\begin{equation}
%\nabla\cdot\Bg\=0\;.
%\end{equation}
%%
Then one also has
\begin{equation}
%\begin{split}
\nabla\cdot\Eg\=\dmup[i]E_i\=-\dmup[i]\frac{\Gscr[0][0][i]}{2}
              \=-8\pi\GN\;\frac{T_{00}}{2}
              \=4\pi\GN\:\rhog\;,
%\end{split}
\end{equation}
using eq.\ \eqref{eq:Einstein_dG} and having defined $\rhog\equiv-T_{00}$\,.\par\smallskip
If we consider the curl of $\Eg$, we obtain
\begin{equation}
\begin{split}
\nabla\times\Eg&\={\varepsilon_i}^{jk}\,\dm[j]E_k
               \=-{\varepsilon_i}^{jk}\,\dm[j]\frac{\Gscr[0][0][k]}{2}
               \=-\frac12\,{\varepsilon_i}^{jk}\,\dm[j]\dm[{[}0]\bhmetr[k{]}][0]\=\\[2\jot]
              &\=-\frac14\,4\;\dm[0]\,{\varepsilon_i}^{jk}\,\dm[j]A_k
               \=-\dm[0]B_i\=-\frac{\dd\Bg}{\dd t}\:.
\end{split}
\end{equation}
Finally, one finds for the curl of $\Bg$
\begin{equation}\label{eq:gravMaxwell4}
\begin{split}
\nabla\times\Bg&\={\varepsilon_i}^{jk}\,\dm[j]B_k
      \=\frac14\,{\varepsilon_i}^{jk}\,
                {\varepsilon_k}^{\ell m}\,\dm[j]\Gscr[0][\ell][m]
      \=\frac14\left({\delta_i}^\ell\delta^{jm}-{\delta_i}^m\delta^{j\ell}\right)\dm[j]\Gscr[0][\ell][m]\=
\\[3\jot]
    &\=\frac12\,\dmup[j]\Gscr[0][i][j]
     \=\frac12\left(\dmup\Gscr[0][i][\mu]+\dm[0]\Gscr[0][i][0]\right)
     \=\frac12\left(\dmup\Gscr[0][i][\mu]-\dm[0]\Gscr[0][0][i]\right)\=
\\[3\jot]
    &\=\frac12\left(8\pi\GN\;T_{0i}-\dm[0]\Gscr[0][0][i]\right)
     \=4\pi\GN\;j_i+\frac{\dd E_i}{\dd t}
     \=4\pi\GN\;\jg\+\frac{\dd\Eg}{\dd t}\:,
\end{split}
\end{equation}
using again eq.\ \eqref{eq:Einstein_dG} and having defined
$\jg \equiv j_i \equiv T_{0i}$\,.\par\medskip
Summarizing, once defined the fields of \eqref{eq:fields0} and having restored physical units, one gets the field equations:
\begin{equation} \label{eq:gravMaxwell}
\begin{split}
&\nabla\cdot\Eg\=4\pi\GN\;\rhog\:,\\[2\jot]
&\nabla\cdot\Bg\=0 \:,\\[2\jot]
&\nabla\times\Eg~=-\dfrac{\dd\Bg}{\dd t} \:,\\[2\jot]
&\nabla\times\Bg\=\frac{4\pi\GN}{c^2}\;\jg
                  \+\frac{1}{c^2}\,\frac{\dd\Eg}{\dd t}\:,
\end{split}
\end{equation}
formally equivalent to Maxwell equations, where $\Eg$ and $\Bg$ are the gravitoelectric and gravitomagnetic field, respectively.
For instance, on the Earth's surface, $\Eg$ corresponds to the Newtonian gravitational acceleration while $\Bg$ is related to angular momentum interactions \cite{Braginsky:1976rb,huei1983calculation,peng1990new,agop2000local}.
The mass current density vector $\jg$ can also be expressed as:
\begin{equation}
\jg \= \rhog\,\mathbf{v}\:,
\end{equation}
where $\mathbf{v}$ is the velocity and $\rhog$ is the mass density.
\par\medskip

\paragraph{Gravito-Lorentz force.}
Let us consider the geodesic equation for a particle in the presence of a weak gravitational field:
\begin{equation}
\frac{d^2x^\lambda}{ds^2}\+\Gam\,\frac{dx^\mu}{ds}\,\frac{dx^\nu}{ds}\=0\:.
\end{equation}
If we consider a non-relativistic motion, the velocity of the particle can be expressed as $\frac{v_i}{c}\simeq\frac{dx^i}{dt}$. If we also neglect terms in the form $\frac{v_i\,v^j}{c^2}$ and limit ourselves to static metric configurations, we find that a geodesic equation for the particle in non-relativistic motion is written as \cite{ruggiero2002gravitomagnetic,mashhoon1989detection}:
\begin{equation}
\frac{d\mathbf{v}}{dt} \= \Eg+\mathbf{v}\times\Bg\:,
\end{equation}
which shows that a free falling particle is governed by the analogous of a Lorentz force produced by the gravito-Maxwell fields.
\par\medskip

\paragraph{Generalized Maxwell equations.}
It is now straightforward to define generalized electric/magnetic fields, scalar and vector potentials, containing both electromagnetic and gravitational contributions, as:
\begin{equation} \label{eq:genpot}
\E=\Ee+\frac{m}{e}\,\Eg\,;\quad\;
\B=\Be+\frac{m}{e}\,\Bg\,;\quad\;
\phi=\phi_\textrm{e}+\frac{m}{e}\,\phig\,;\quad\;
\A=\Ae+\frac{m}{e}\,\Ag\,,
\end{equation}
where $m$ and $e$ are the electron mass and charge, respectively.\par\smallskip
The generalized Maxwell equations then become:
\begin{equation} \label{eq:genMaxwell}
\begin{split}
&\nabla\cdot\E\=\left(\frac1\epsg+\frac{1}{\epsz}\right)\,\rho \:,\\[2\jot]
&\nabla\cdot\B\=0 \:,\\[2\jot]
&\nabla\times\E\:=-\dfrac{\dd\B}{\dd t} \:,\\[2\jot]
&\nabla\times\B\=\left(\mug+\muz\right)\,\jj
                  \+\frac{1}{c^2}\,\dfrac{\dd\E}{\dd t} \:,
\end{split}
\end{equation}
where $\epsz$ and $\muz$ are the electric permittivity and magnetic permeability in the vacuum, and where we have set
\begin{equation}
\rhog\=\frac{m}{e}\,\rho\:,\qquad\quad
\jg\=\frac{m}{e}\,\jj\:,
\end{equation}
$\rho$ and $\jj$ being the electric charge density and electric current density, respectively. The introduced vacuum gravitational permittivity $\epsg$ and vacuum gravitational permeability $\mug$ are defined as
\begin{equation}
\epsg=\frac{1}{4\pi\GN}\,\frac{e^2}{m^2}\:,\qquad\quad
\mug=\frac{4\pi\GN}{c^2}\,\frac{m^2}{e^2}\:.
\end{equation}
\par\bigskip
In this Section we have then shown how to define a new set of generalized Maxwell equations for generalized electric $\E$ and magnetic $\B$ fields, in the limit of weak gravitational field. In the following, we are going to use these results to analyse the interaction between a superconducting sample and the weak, static Earth's gravitational field.

%%%%%%%%%%%%%%%%%%%%%%%%%%%%%%%%%%%%%%%%%%%%%%%%

\section{The model}
Now we are going to study in detail the conjectured gravity/superconductivity interplay
%between a suitable superconducting sample and the Earth's static %gravitational field,
making use of the Ginzburg--Landau formulation combined with the described gravito-Maxwell formalism. In particular, we write the Ginzburg--Landau equations for a superconducting sample in the weak, static Earth's gravitational field. The latter is formally treated as the gravitational component of a generalized electric field, exploiting the formal analogy discussed in the previous Section \ref{sec:gravMax}.

\subsection{Time-dependent Ginzburg--Landau formulation}
Since the gravitoelectric field is formally analogous to a generalized electric field, we can use the time-dependent Ginzburg--Landau equations (TDGL) written in the form \cite{ullah1991effect,tang1995time,du1996high,lin1997ginzburg,fleckinger1998dynamics,
kopnin1999time,ghinovker1999explosive}:
\begin{subeqs} \label{eq:TDGL0}
\begin{align}
&\frac{\hbar^2}{2\,m}\left(i\,\nabla
        +\frac{2\,e}{\hbar}\,\A\right)^{\!2}\psi
      \,-\,a\,\psi\,+\,b\,\abs{\psi}^2\psi
    \:=\,-\,\frac{\hbar^2}{2\,m\,\mathcal{D}}\left(\frac{\dd}{\dd t}
      \,+\,\frac{2\,i\,e}{\hbar}\,\phi\right)\,\psi\;,
\label{subeq:TDGL1}
\\[2\jot]
&\nabla\times\nabla\times\A\,-\,\nabla\times\B
    \=\mu_0\,\big(\jj_\text{n}+\jj_\text{s}\big)\,,
\end{align}
\end{subeqs}%   always put % here!
where $\jj_\text{n}$ and $\jj_\text{s}$ are expressed as
\begin{equation}
\begin{split}
\jj_\text{n}&\:=-\,\sigma\left(\frac{\dd\A}{\dd t}+\nabla\phi\right)\:,
\\[3\jot]
\jj_\text{s}&\:=-\,i\,\hbar\,\frac{e}{m}\left(\psi^*\,\nabla\psi-\psi\,\nabla\psi^*\right)
    -\frac{4\,e^2}{m}\,\abs{\psi}^2\A\:.
\end{split}
\end{equation}
and denote the contributions related to the normal current and supercurrent densities, respectively.%
\footnote{The TDGL equations \eqref{eq:TDGL0} for the variables $\psi$, $\A$ are derived minimizing the total Gibbs free energy of the system \cite{tinkham2004introduction,ketterson1999superconductivity,
DeGennes2018superconductivity}.
}
In the above expressions, $\mathcal{D}$ is the diffusion coefficient, $\sigma$ is the conductivity in the normal phase, $\B$ is the applied magnetic field and the vector potential $\A$ is minimally coupled to $\psi$.
The coefficients $a$ and $b$ in \eqref{subeq:TDGL1} have the following form:
\begin{equation}
a\=a(T)\=a_{0}\,(T-\Tc)\,,\qquad\qquad
b\=b(\Tc)\,,\qquad
\end{equation}
$a_0$, $b$ being positive constants and $\Tc$ the critical temperature of the superconductor. The boundary and initial conditions are
%
%\begingroup%
%\setlength{\belowdisplayskip}{7pt plus 3pt minus 4pt}%
\begin{align}
  \left.
  \begin{aligned}
  \left(i\,\nabla\psi+\frac{2\,e}{\hbar}\,\A\,\psi\right)\cdot\mathbf{n}=0&  \cr
  \hfill\nabla\times\A\cdot\mathbf{n}=\B\cdot\mathbf{n}&  \\[1.5\jot]
  \hfill\A\cdot\mathbf{n}=0&
  \end{aligned}
  \;\;\right\} \; \text{on }\dd\Omega\times(0,t)\;,
\qquad\quad
  \left.
  \begin{aligned}
  \psi(x,0)&\=\psi_0(x) \cr
  \A(x,0)&\=\A_0(x)
  \end{aligned}
  \!\!\!\!\right\} \; \text{on }\Omega\;,\qquad
\label{eq:boundary}
\end{align}
%\endgroup
%
where $\dd\Omega$ is the boundary of a smooth and simply connected domain in $\mathbb{R}^\ms{\textrm{N}}$.

\paragraph{Dimensionless TDGL.}
In order to write eqs.\ \eqref{eq:TDGL0} in a dimensionless form, the following expressions can be introduced:
\begin{equation} \label{eq:param}
\begin{split}
&\Psi^2(T)=\frac{\abs{a(T)}}{b}\,,\qquad\;
\xi(T)=\frac{\hbar}{\sqrt{2\,m\,|a(T)|}}\,,\qquad\;
\lambda(T)=\sqrt{\frac{b\,m}{4\,\mu_0\,|a(T)|\,e^2}}\,,\qquad\;
\kappa=\frac{\lambda(T)}{\xi(T)}\,,\quad
\\[2\jot]
&\tau(T)=\frac{\lambda^2(T)}{\mathcal{D}}\,,\qquad\;
\eta=\mu_0\,\sigma\,\mathcal{D}\,,\qquad\;\;
\Bc(T)=\sqrt{\frac{\mu_0\,\abs{a(T)}^2}{b}}=
    \frac{\hbar}{2\sqrt{2}\,e\,\lambda(T)\,\xi(T)}\,,
\end{split}
\end{equation}
where $\lambda(T)$, $\xi(T)$ and $\Bc(T)$ are the penetration depth, coherence length and thermodynamic critical field, respectively. We also define the dimensionless quantities
\begin{equation}
t'=\:\frac{t}{\tau}\,,\qquad\;
x'=\:\frac{x}{\lambda}\,,\qquad\;
y'=\:\frac{y}{\lambda}\,,\qquad\;
%z'=\:\frac{z}{\lambda}\,,\qquad\;
\psi'=\:\frac{\psi}{\Psi}\,,\qquad
\label{eq:dimlesscoord}
\end{equation}
and the new dimensionless fields and currents
\begin{equation}
\A'=\frac{\A\,\kappa}{\sqrt{2}\,\Bc\,\lambda}\,,\qquad
\phi'=\frac{\phi\,\kappa}{\sqrt{2}\,\Bc\,\mathcal{D}}\,,\qquad
\E'=\frac{\E\,\lambda\,\kappa}{\sqrt{2}\,\Bc\,\mathcal{D}}\,,\qquad
\B'=\frac{\B\,\kappa}{\sqrt{2}\,\Bc}\,,\qquad
\jj'=\frac{\jj\,\mu_0\,\lambda\,\kappa}{\sqrt{2}\,\Bc}\,.
\label{eq:dimlessfields}
\end{equation}
Inserting \cref{eq:dimlesscoord,eq:dimlessfields} in eqs.\ \eqref{eq:TDGL0} and dropping the primes, gives the dimensionless TDGL equations in a bounded, smooth and simply connected domain in $\mathbb{R}^\ms{\textrm{N}}$ \cite{tang1995time,lin1997ginzburg}:
\begin{subeqs}\label{eq:dimlessTDGL}
\begin{align}
&\frac{\dd\psi}{\dd t} \,+\, i\,\phi\,\psi
    \,+\,\kappa^2\left(\abs{\psi}^2-1\right)\,\psi
    \,+\,\left(i\,\nabla+\A\right)^2 \psi\=0 \:,
\label{subeq:dimlessTDGLn1}
\\[2\jot]
&\nabla\times\nabla\times\A\,-\,\nabla\times\B
    \=\jj_\text{n}+\jj_\text{s}
    \:=\,-\,\eta\,\left(\frac{\dd\A}{\dd t}+\nabla\phi\right)
    -\frac{i}{2}\left(\psi^*\nabla\psi-\psi\nabla\psi^*\right)
    -\abs{\psi}^2\A\:,
\label{subeq:dimlessTDGLn2}
\end{align}
\end{subeqs}%   always put % here!
and the boundary and initial conditions \eqref{eq:boundary} become, in the dimensionless form
\begin{align}
  \left.
  \begin{aligned}
  \left(i\,\nabla\psi+\A\,\psi\right)\cdot\mathbf{n}=0&\cr
  \nabla\times\A\cdot\mathbf{n}=\B\cdot\mathbf{n}&\cr
  \A\cdot\mathbf{n}=0&
  \end{aligned}
  \!\!\!\right\} \; \text{on }\dd\Omega\times(0,t)\;;
\qquad\quad
  \left.
  \begin{aligned}
  \psi(x,0)&\=\psi_0(x)\cr
  \A(x,0)&\=\A_0(x)
  \end{aligned}
  \!\!\!\!\right\} \; \text{on }\Omega\;.\qquad
\label{eq:dimlessboundary}
\end{align}

\subsection{Solving dimensionless TDGL}
We now study the possible local alterations of the Earth's gravitational field (weak uniform field) inside a superconductor.
Let us consider the dimensionless form of the time-dependent Ginzburg--Landau equations in the gauge of vanishing scalar potential $\phi=0$ \,\cite{gulian2020shortcut}:%
\footnote{%
Here we decide to use the most convenient option for subsequent calculations, since any gauge choice shall not influence any physical results, being the equations gauge-invariant.
From a physical point of view, the choice is also motivated by the fact that there are no localized charges in the superconductor, while any contribution to the total gravitational field coming from the superconductor mass is irrelevant and can be neglected}
\begin{subeqs}\label{eq:TDGL}
\begin{align}
\frac{\partial\psi}{\partial t}\:&=\,
    -\left(i\,\nabla+\A\right)^{2}\psi\-\kappa^2\left(\abs{\psi}^2-1\right)\psi\:,
\label{subeq:TDGL1}
\\[2\jot]
\eta\,\frac{\partial\A}{\partial t}\:&=
    \,-\,\nabla\times\nabla\times\A\,+\,\nabla\times\B
    \,-\,\frac{i}{2}\left(\psi^*\nabla\psi-\psi\nabla\psi^*\right)
    \,-\,\abs{\psi}^2\A\:,
\label{subeq:TDGL2}
\end{align}
\end{subeqs}%   always put % here!
where $\psi\equiv\psi(\x,t)$ is a complex function that we express as
\begin{equation}
\psi\=\abs{\psi}\,\exp(i\,\theta)\=\Real\,\psi+i\,\Img\,\psi\=\psi_1+i\,\psi_2\:,
\end{equation}
so that \eqref{subeq:TDGL1} gives two distinct equations for the real and imaginary parts $\psi_1$ and $\psi_2$.\par

\paragraph{1-D case.}%in the $\phi=0$ gauge
Let us now restrict to the 1-dimensional case \big(\,$\nabla\rightarrow\partial/\partial x$, \;$\A\rightarrow A_x\equiv A$\,\big). In this situation, the above TDGL \eqref{eq:TDGL} give rise to the following equations:
\begin{equation}
\begin{split}
\dt[\psi_1]&\=
    \ddx[\psi_1]+A\,\dx[\psi_2]+\psi_2\,\dx[A]-\psi_1\,A^2
    -\kappa^2\left(\abs{\psi_1}^2+\abs{\psi_2}^2-1\right)\psi_1\:,
\\[2\jot]
\dt[\psi_2]&\=
    \ddx[\psi_2]-A\,\dx[\psi_1]-\psi_1\,\dx[A]-\psi_2\,A^2
    -\kappa^2\left(\abs{\psi_1}^2+\abs{\psi_2}^2-1\right)\psi_2\:,
\\[2\jot]
\eta\,\dt[A]&\=-\left(\psi_2\,\dx[\psi_1]-\psi_1\,\dx[\psi_2]\right)
    -\left(\psi_1^2+\psi_2^2\right)A\:,
\end{split}
\end{equation}
since, in one dimension, \,$\nabla^2 A\,=\,\tfrac{\partial}{\partial x}\,\left(\nabla\cdot\A\right)$\, and then
\begin{equation}
\nabla\times\nabla\times\A\;=\;\nabla\,\left(\nabla\cdot\A\right)-\nabla^2 A
    \overset{\tts{1d}}{\;=\;}0\:.
\end{equation}
Now, we consider a half-infinite superconductive region, where the $\vec{x}$ direction is perpendicular to superconductor surface (coinciding with the $yz$ plane), i.e.\ we imagine that for $x>0$ we have an empty space, while the region occupied by the material is located at $x\leq0$. The system is immersed in a static, uniform gravitational field \,$\Eg^{\textsc{ext}}=-g\,\ux$\,, where $g$ is the standard gravity acceleration. We are in the gauge where, in the \emph{dimensional} form, we can write for the gravitoelectric field inside the superconductor
\begin{equation}
\Eg=-\dt[\A_\text{g}(t)]\:,
\end{equation}
while the external gravitational vector potential outside the superconductor is given by
\begin{equation}
\A_\text{g}^{\!\textsc{ext}}(t)=g\left(C+t\right)\,\ux\:,
\end{equation}
where $C$ is a constant. In the 1-D \emph{dimensionless} form, dropping the primes, we have
\begin{equation}
A^{\textsc{ext}}\=\frac{m}{e}\,A_\text{g}^{\textsc{ext}}\,\frac{\kappa}{\sqrt{2}\,\Bc\,\lambda}\=
    \gstar\left(c_1+t\right)\:,
\end{equation}
with
\begin{equation}
c_1\,=\,\cfrac{C}{\tau}\;,\qquad\quad\;\;
\gstar\=\frac{m\,\kappa\,\lambda(T)\,g}{\sqrt{2}\,e\,\mathcal{D}\,\Bc(T)}\:\ll\:1\:.
\end{equation}
having used relations \eqref{eq:param}.\par
Next, we express the $\psi_1$, $\psi_2$ and $A$ fields as:
\begin{subeqs}
\begin{align}
\psi_1(x,t)&\=\psi_{10}(x)+\gstar\,\gamma_1(x,t)\:,
\\[1.5\jot]
\psi_2(x,t)&\=\psi_{20}(x)+\gstar\,\gamma_2(x,t)\:,
\\[1.5\jot]
A(x,t)&\=\gstar\,\beta(x,t)\:,\label{subeq:A_gstBeta}
\end{align}
\end{subeqs}% <-------
where $\psi_{10}$ and $\psi_{20}$ represent the unperturbed system and satisfy
\begin{subeqs}
\begin{align}
0&\=\frac{1}{\kappa^2}\,\ddx[\psi_{10}]+\psi_{10}
    -\psi_{10}\left(\psi_{10}^2+\psi_{20}^2\right)\:,
\label{subeq:psi10}
\\[1.5\jot]
0&\=\frac{1}{\kappa^2}\,\ddx[\psi_{20}]+\psi_{20}
    -\psi_{20}\left(\psi_{10}^2+\psi_{20}^2\right)\:.
\end{align}
\end{subeqs}% <- always put % here!
The $\psi_{10}$ and $\psi_{20}$ components satisfy the same kind of equation, and we choose to set $\psi_{20}=0$ (also implying $\psi_{0}=\psi_{10}+i\,\psi_{20}=\psi_{10}\,\in\mathbb{R}$), so that $\psi_{10}=\tanh\left(\tfrac{\kappa x}{\sqrt{2}}\right)$ gives the standard solution for \eqref{subeq:psi10} \cite{ketterson1999superconductivity}.
%giving for the $\psi_{10}$ component
%%
%\begin{equation}
%0\=\frac{1}{\kappa^2}\,\ddx[\psi_{10}]+\psi_{10}-\psi_{10}^3\:.
%\end{equation}
%
We are then left with the following set of equations:
\begin{subeqs}\label{eq:TDGLgam1gam2beta}
\begin{align}
%\begin{equation}
%\begin{split}
\dt[\gamma_1]&\=
    \ddx[\gamma_1]+\kappa^2\left(1-3\,\psi_{10}^2\right)\gamma_1\:,
\\[2\jot]
\dt[\gamma_2]&\=
    \ddx[\gamma_2]+\kappa^2\left(1-3\,\psi_{10}^2\right)\gamma_2
    -\beta\,\dx[\psi_{10}]-\psi_{10}\,\dx[\beta]\:,
\\[2\jot]
\eta\,\dt[\beta]&\=
    -\gamma_2\,\dx[\psi_{10}]+\psi_{10}\,\dx[\gamma_2]-\psi_{10}^2\,\beta\:,
\label{subeq:TDGLbeta}
%\end{split}
%\end{equation}
\end{align}
\end{subeqs}%   always put % here!
where the last \eqref{subeq:TDGLbeta} implies that $\beta(x,t)$ does not depend on $\gamma_1(x,t)$.
If we decide to put ourselves away from borders, we can set $\psi_{10}\simeq1$ in equations \eqref{eq:TDGLgam1gam2beta}, obtaining
\begin{subeqs}
\begin{align}
\dt[\gamma_1]&\:\simeq\:
    \ddx[\gamma_1]-2\,\kappa^2 \gamma_1\:,
\\[2\jot]
\dt[\gamma_2]&\:\simeq\:
    \ddx[\gamma_2]-2\,\kappa^2 \gamma_2-\dx[\beta]\:,
\\[2\jot]
\eta\,\dt[\beta]&\:\simeq\:
    \dx[\gamma_2]-\beta\:,
\end{align}
\end{subeqs}%   always put % here!
that gives for $\beta$ the explicit solution
\begin{equation}
\beta(x,t)\=e^{-\tfrac{t}{\eta}}\,\left(b_1(x)
    +\frac{1}{\eta}\;\int^t_0\!\!dt\;e^{\tfrac{t}{\eta}}\;
    \dx[\gamma_2(x,t)]\right)\:.
\end{equation}
where $b_1(x)=c_1$, as it is implied by eq.\ \eqref{subeq:A_gstBeta} for $t\simeq0$.\par\smallskip
Let us keep in mind that we are considering a semi-infinite superconductor whose surface is parallel to the ground and normal to the $\vec{x}$ axis (one-dimensional case) where the external vector potential is expressed as:
\begin{equation}
A^{\textsc{ext}}(t)\=\left(c_1+t\right)\,\gstar\;.
\end{equation}
At the time $t=0$, the sample goes in the superconductive state, while we make the natural assumption that in the normal state ($t<0$) the material has just the standard (Newtonian) interaction with the Earth's gravity, implying that the local gravitational field assumes the same values inside and outside the sample for $t<0$. We then write the following boundary conditions:
\begin{equation}\label{eq:boundarycond}
\begin{split}
\begin{alignedat}{3}
&\psi(0,t)=0\,,\qquad\quad
&&\psi(x,0)=\psi_{10}(x)\,,\qquad\quad
&&\dx[\psi_{1}](x,0)=0\,,
\\[\jot]
&\gamma_{1}(0,t)=0\,,\qquad\quad
&&\gamma_{1}(x,0)=0\,,\qquad\quad
&&\dx[\gamma_{1}](x,0)=0\,,
\\[\jot]
&\gamma_{2}(0,t)=0\,,\qquad\quad
&&\gamma_{2}(x,0)=0\,,\qquad\quad
&&\dx[\gamma_{2}](x,0)=0\,,
\end{alignedat}
\end{split}
\end{equation}
together with the condition
\begin{equation}
\lim_{\;t\to 0}\,\gstar\,\dt[\beta](x,t)\=\gstar\:.
\end{equation}
implying that the effect takes place when the superconducting phase appears.\par
Let us now fix the constant $c_1$. Using \eqref{subeq:TDGLbeta}, we can express the relation between $E_\textrm{g}$ and $\beta$ as
\begin{equation}
\frac{E_\textrm{g}}{\gstar}\=-\dt[\beta]\=
    \frac{1}{\eta}\left(\gamma_2\,\dx[\psi_{10}]-\psi_{10}\,\dx[\gamma_2]\right)
    +\frac{\psi_{10}^2}{\eta}\,\beta\:.
\end{equation}
%
%
%The last conditions originates from the natural hypothesis that the effect on %gravitational field only exists when the material goes in the superconductive %state, i.e.\ for $t>0$.
Given the natural hypothesis that the affection of the gravitational field only exists when the material is in the superconductive state ($t>0$), we expect that, at initial time,
\begin{equation}
\lim_{\;t\to 0^+}\,\frac{E_\textrm{g}}{\gstar}\=1\:,
\end{equation}
while from conditions \eqref{eq:boundarycond} we also have
\begin{equation}
\lim_{\;t\to 0^+}\,\gamma_2(x,t)\=0\,,\qquad\quad
\lim_{\;t\to 0^+}\,\dx[\gamma_2](x,t)\=0\,,
\end{equation}
from which we get in turn
\begin{equation}
1\=\frac{\psi_{10}^2}{\eta}\:\beta(x,0^+)\=
    \frac{\psi_{10}^2}{\eta}\:\frac{A^{\textsc{ext}}(0^+)}{\gstar}\=
    \frac{\psi_{10}^2}{\eta}\:c_1
\;\quad\Longrightarrow\quad\;
c_1=\frac{\eta}{\psi_{10}^2}\:.
\end{equation}
This constant is ineffective in the empty space, while it determines physical effects in the superconductive state.
The above formulation shows how the described interplay should work:
the external gravitational field is affected by the presence of the sample only when it goes in the superconductive state (when the vector potential starts to ``feel'' the presence of the superfluid). From the other side, the external gravitational vector potential seems involved in the material superconductive transition, since the external constant $c_1$ tends to assume a fixed value related to the properties of the superfluid entering the superconducting state.\par
Now we can rewrite the explicit solution for $\beta(x,t)$ away from borders $(\psi_{10}\simeq1)$:
\begin{equation}
\beta(x,t)\=e^{-\tfrac{t}{\eta}}\,\left(\eta
    +\frac{1}{\eta}\;\int^t_0\!\!dt\;e^{\tfrac{t}{\eta}}\;
    \dx[\gamma_2(x,t)]\right)\:,
\end{equation}
from which we get the ratio
\begin{equation}
\frac{\Eg}{\gstar}\=-\frac{\partial \beta(x,t)}{\partial t}\=
    \frac{1}{\eta}\,e^{-\tfrac{t}{\eta}}\,\left(\eta
    +\frac{1}{\eta}\;\int^t_0\!\!dt\;e^{\tfrac{t}{\eta}}\;
    \dx[\gamma_2(x,t)]\right)-\frac{1}{\eta}\frac{\partial \gamma_2(x,t)}{\partial x}\:.
\label{eq:ratio_Eg_g}
\end{equation}

\section{Discussion}
Given the explicit expression \eqref{eq:ratio_Eg_g} for the ratio $\Eg/\gstar$, we can estimate, for $t \simeq 0^{+}$, the value of gravitational field inside the superconductor:
\begin{equation}
t \simeq 0^{+}\::\qquad
    \frac{\Eg}{\gstar}=1-\frac{t}{\eta}-\frac{1}{\eta}\frac{\partial \gamma_2(x,0^+)}{\partial x}\:.
\label{eq:Eg_g_t0}
\end{equation}
In the superconductive state, the gravitational field is modified in a way that depends on physical characteristic of the particular material. We can see from the above \eqref{eq:Eg_g_t0} that the involved quantities are $\eta$ and the spatial derivative of $\gamma_2$.\par\smallskip
Let us discuss which should be the most favourable choices for the parameters to enhance the desired interaction. First of all, we would like to maximize $\tfrac{\partial\gamma_2}{\partial x}$: to do this, it is sufficient to introduce disorder in the material, induced, for instance, by means of proton
irradiation or chemical doping.
Then, we also want a small $\eta$ parameter: being the latter proportional to the product of the diffusion coefficient times the conductivity just above $\Tc$, it is necessary to have materials that in the normal state are bad conductors and have low Fermi energies, such as cuprates.\par
It is also very important to maximize the time scale ($\tau=\lambda^2/\mathcal{D}$) in order to better observe the effect. This is achieved by increasing the penetration length and reducing the diffusivity coefficient, just as it occurs in superconducting cuprates with disorder.\par
In Tables \ref{tab:YBCOvsPb1} and \ref{tab:YBCOvsPb2} it is possible to see typical parameters of low (Pb) and high (YBCO) $\Tc$ superconductors, some of which
calculated at a temperature $T_{\star}$ such that the quantity $\tfrac{\Tc-T_{\star}}{\Tc}$ is the same in the two materials.
If we go closer to $\Tc$, it is possible to increase the effect:
for example, in the case of YBCO, at $T=87\,\Kelv$ the $\tau$ parameter is of the order of $10^{-9}\,\mathrm{s}$ and the reduction of the gravitational field is of the order of $10^{-7}$, having neglected the last term in eq.\ \eqref{eq:Eg_g_t0} (in high--$\Tc$ superconductors not irradiated, we usually have low disorder, so that the spatial derivative of $\gamma_2$ is small).
%
%It will be interesting to do the same calculations at $T\lesssim\Tc$ in presence of a external uniform magnetic field
%$H\lesssim\Hctwo$ and a external uniform electric field parallel to superconductor plane surface and in presence of defects near to critical temperature.

\section{Concluding remarks}
We have shown how the gravito-Maxwell formalism can be instrumental in describing a gravity/superfluid interplay, when combined with the condensed matter formalism of the time-dependent Ginzburg--Landau equations. Our analysis suggests that a non-negligible interaction could be present, despite the experimental detection difficulties that may arise, especially in relation to the short time intervals in which the effect occurs. In particular, the dimensionless TDGL can provide qualitative and quantitative suggestion about the magnitude of the interaction, once chosen appropriate boundary conditions.\par
Clearly, proper arrangement of the experimental setup is crucial to maximize the effect. In particular, the focus should be on suitable sample geometry, material parameters and laboratory settings, so as to enhance the interaction in workable time scales \cite{Ummarino:2019cvw,Gallerati:2020tyq,Ummarino:2020loo}. It is also possible that a significant improvement comes from the presence of external electric and magnetic fields, since the latter determine the presence of moving vortices, giving rise to a possible additional affection of the local gravitational field.

\begin{table}[!htp]
%\captionsetup{aboveskip=20pt}
\centering
\makegapedcells
\setcellgapes{5pt} % vertical space for cells
\begin{tabular}%{@{}ccc@{}}
{@{} M{p}{0.15} M{p}{0.175} M{p}{0.25} @{}}
% @{} prevents white spaces at ends of table
\toprule
\midrule
                   &  \text{YBCO}                 &  \text{Pb}  \\
\thinrule
\Tc                &   89\,\Kelv                  &  7.2\,\Kelv \\
T_\star            &   77\,\Kelv                  &  6.3\,\Kelv \\
\xi(T_\star)       &   3.6\cdot10^{-9}~\mt        &  1.7\cdot10^{-7}~\mt \\
\lambda(T_\star)   &   3.3\cdot10^{-7}~\mt        &  7.8\cdot10^{-8}~\mt \\
\sigma^{-1}        &   4.0\cdot10^{-7}~\Omega\,\mt\,{}^\ms{(\ast)}
                                                  &
                                           2.5\cdot10^{-9}~\Omega\,\mt\,{}^\ms{(\ast\ast)}\\
\Bc(T_\star)       &   0.2~\mathrm{Tesla}         &  0.018~\mathrm{Tesla}\\
\kappa             &   94.4                       &  0.48 \\
\tau(T_\star)      &   3.4\cdot10^{-10}~\s        &  6.1\cdot10^{-15}~\s \\
\eta               &   1.3\cdot10^{-2}            &  6.6\cdot10^{3} \\
\gstar             &   2.0\cdot10^{-11}           &  8.2\cdot10^{-17} \\
\mathcal{D}        &   3.2\cdot10^{-4}~\mt^{2}/\s &  1~\mt^{2}/\s \\
\ell               &   6.0\cdot10^{-9}~\mt        &  1.7\cdot10^{-6}~\mt \\
\vF                &   1.6\cdot10^{5}~\mt/\s      &  1.8\cdot10^{6}~\mt/\s \\
{}                 &  \text{\fns${}^\ms{(\ast)}\;T\,=\,90\,\Kelv$}
                                                  &\text{\fns${}^\ms{(\ast\ast)}\;T\,=\,15\,\Kelv$}\\
\midrule
\bottomrule
\end{tabular}
%\smallskip
\caption{YBCO vs.\ Pb.}
\label{tab:YBCOvsPb1}
\end{table}
%\parbox{0.3\textwidth}{\centering$4\cdot10^{-7}~\Omega\mt$\\($T=90\,\Kelv$)}

\bigskip

\begin{table}[!htp]
\small
\noindent
\centering
\makegapedcells
\setcellgapes{3pt}
\begin{adjustwidth}{-3cm}{-3cm}% change margins
\begin{center}
\begin{subtable}[t]{0.7\textwidth}
\centering
\begin{tabular}%{@{}ccc@{}}
{@{} M{p}{0.15} M{p}{0.17} M{p}{0.2} M{p}{0.15} @{}}
% @{} prevents white spaces at ends of table
\toprule
\midrule
\textbf{YBCO}  &  \lambda      &  \tau                 & \gstar \\
\thinrule
T=0\,\Kelv\hphantom{0}%
        & 1.7\cdot10^{-7}~\mt  &  9.03\cdot10^{-11}~\s  & 2.6\cdot10^{-12}
\\
T=70\,\Kelv & 2.6\cdot10^{-7}~\mt & 2.1\cdot10^{-10}~\s & 9.8\cdot10^{-12}
\\
T=77\,\Kelv & 3.3\cdot10^{-7}~\mt & 3.4\cdot10^{-10}~\s & 2\cdot10^{-11}
\\
T=87\,\Kelv & 8\cdot10^{-7}~\mt   &  2\cdot10^{-9}~\s   & 2.8\cdot10^{-7}
\\
\midrule
\bottomrule
\end{tabular}
%\smallskip
%\caption{YBCO}
\label{subtab:YBCO}
\end{subtable}% <------
\hspace{-2.25em}
\begin{subtable}[t]{0.7\textwidth}
\centering
\begin{tabular}%{@{}ccc@{}}
{@{} M{p}{0.18} M{p}{0.17} M{p}{0.2} M{p}{0.15} @{}}
% @{} prevents white spaces at ends of table
\toprule
\midrule
\textbf{Pb} &  \lambda            &  \tau                & \gstar \\
\thinrule
T=0\,\Kelv\hphantom{.00}%
          & 3.90\cdot10^{-8}~\mt& 1.5\cdot10^{-15}~\s & 1\cdot10^{-17}
          \\
T=4.20\,\Kelv & 4.3\cdot10^{-8}~\mt & 1.8\cdot10^{-15}~\s & 1.4\cdot10^{-17}
\\
T=6.26\,\Kelv & 7.8\cdot10^{-8}~\mt & 6.1\cdot10^{-15}~\s & 8.2\cdot10^{-17}
\\
T=7.10\,\Kelv & 2.3\cdot10^{-7}~\mt & 5.3\cdot10^{-14}~\s & 2.2\cdot10^{-15}
\\
\midrule
\bottomrule
\end{tabular}
%\smallskip
%\caption{Pb}
\label{subtab:Pb}
\end{subtable}% <---------
%
%\smallskip
\end{center}
\end{adjustwidth}
\caption{YBCO and Pb parameters at different temperatures.}
\label{tab:YBCOvsPb2}
\end{table}

%%%%%%%%%%%%%%%%%%%%%%%%%%%%%%%%

\bigskip
\section*{\normalsize Acknowledgments}
\vspace{-5pt}
This work was supported by the MEPhI Academic Excellence Project (contract No.\ 02.a03.21.0005) for the contribution of prof.\ G.\ A.\ Ummarino.
We also thank Fondazione CRT \,\includegraphics[height=\fontcharht\font`\B]{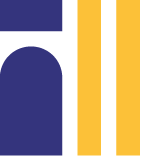}\:
that partially supported this work for dott.\ A.\ Gallerati.

%---- APPENDICES ------------------------------------------
%
%\appendix
%%\phantomsection % use it for correct TOC link !!!
%%\addcontentsline{toc}{section}{Appendices} % add References to TOC
%%\addtocontents{toc}{\protect\setcounter{tocdepth}{0}}%no_single_Appendices_in_TOC
%\addtocontents{toc}{\protect\addvspace{2.5pt}}%
%\numberwithin{equation}{section}%
%%

%---- BIBLIOGRAPHY ------------------------------------------

%\bigskip
\pagebreak
\hypersetup{linkcolor=blue}
\phantomsection % use it for correct TOC link !!!
\addtocontents{toc}{\protect\addvspace{3.5pt}}% add vertical space in TOC
\addcontentsline{toc}{section}{References} % add References to TOC
\bibliographystyle{mybibstyle}
%\small
\bibliography{bibliografia} % The file containing the bibliography

%---------------------------------------

\end{document}